\documentclass[pre,amsmath,amssymb,superscriptaddress]{revtex4}

\usepackage{graphicx}
\usepackage{color}
\usepackage{latexsym}

\definecolor{linkcolor}{rgb}{0,0,0.6} 

\newcommand{\mean}[1]{{\langle {#1} \rangle}}

\begin{document}

\title{Folding Rate Optimization Promotes Frustrated Interactions in Entangled Protein Structures}

\author{Federico Norbiato}
\affiliation{
Department of Physics and Astronomy, University of Padova, 
Via Marzolo 8, I-35131 Padova, Italy
}

\author{Flavio Seno}
\affiliation{
Department of Physics and Astronomy, University of Padova, 
Via Marzolo 8, I-35131 Padova, Italy
}
\affiliation{
INFN, Sezione di Padova, Via Marzolo 8, I-35131 Padova, Italy
}

\author{Antonio Trovato}
\affiliation{
Department of Physics and Astronomy, University of Padova, 
Via Marzolo 8, I-35131 Padova, Italy
}
\affiliation{
INFN, Sezione di Padova, Via Marzolo 8, I-35131 Padova, Italy
}

\author{Marco Baiesi}
\email{baiesi@pd.infn.it}
\affiliation{
Department of Physics and Astronomy, University of Padova, 
Via Marzolo 8, I-35131 Padova, Italy
}
\affiliation{
INFN, Sezione di Padova, Via Marzolo 8, I-35131 Padova, Italy
}
\begin{abstract}
  Many native structures of proteins accomodate complex
  topological motifs such as knots, lassos, and other geometrical
  entanglements. How proteins can fold quickly even in the presence of
  such topological obstacles is a debated question in structural
  biology. Recently, the hypothesis that energetic frustration might
  be a mechanism to avoid topological frustration has been put forward
  based on the empirical observation that loops involved in
  entanglements are stabilized by weak interactions between
  amino-acids at their extrema. To verify this idea, we use a toy
  lattice model for the folding of proteins into two almost identical
  structures, one entangled and one not.  As expected, the folding
  time is longer when random sequences folds into the entangled
  structure. This holds also under an evolutionary pressure simulated
  by optimizing the folding time. It turns out that optmized protein
  sequences in the entangled structure are in fact characterized by
  frustrated interactions at the closures of entangled loops. This
  phenomenon is much less enhanced in the control case where the
  entanglement is not present. Our findings, which are in agreement
  with experimental observations, corroborate the idea that an
  evolutionary pressure shapes the folding funnel to avoid topological
  and kinetic traps.
\end{abstract}
\maketitle


\section{Introduction}

The biological function of most proteins requires them to fold into a
well-defined native state, implying that both structure maintenance
and efficient folding are kept under selective pressure by
evolutionary processes~\cite{sikosek2014}. In particular, a direct
experimental evidence, pointing to some degree of folding rate
optimization throughout evolution, was recently provided for
ribonuclease H, using ancestral sequence
reconstruction~\cite{lim2016}. Bio-informatics analyses had also
uncovered similar evolutionary signals already two decades ago for
several folds~\cite{mirny1999}, and more recently for a large catalog
of protein domains~\cite{debes2013}.

The latter study was based on the well known empirical correlation
between experimentally measured folding rates of proteins and simple
descriptors of the structural organization of the native
state~\cite{ivankov2003}. More general features of the folding
mechanism are as well dictated by the overall topology of the native
state~\cite{baker2000}. In fact, contact order~\cite{plaxco1998} and
other related descriptors are based on the topological properties of
the network formed by pairs of residues that are nearby in the
three-dimensional space~\cite{mugler2014}. The simpler the network,
the faster the predicted folding. The topology of the network of
contacts, however, does not necessarily capture the topology of the
protein backbone seen as a curve in the three-dimensional space, and
the possible formation of knots and other entangled motifs.

The discovery of knots in few proteins~\cite{taylor2000} came indeed
as a surprise because they seem an unnecessary complication for the
folding. Their presence could be related to some biological function
or stability requirement~\cite{jackson2017,dabrowski2017tie}, and the
mechanisms allowing the dynamics to thread the protein backbone to
form knots are under intense
investigation~\cite{virnau2006,lua2006,Bolinger2010,Rawdon2015,Jarmolinska2019}.

After knots, it was realized that other topological motifs may tangle
the three-dimensional structure of some proteins. These include
knotoids~\cite{Goundaroulis2017}, slipknots~\cite{Sulkowska2012},
lassos~\cite{frechet1994,niemyska2016}, pokes~\cite{khatib2009} and
other forms of
entanglement~\cite{Baiesi_et_al_SciRep_2016,Baiesi_et_al_JPA_2017,Zhao2018,Baiesi_SR_2019}
related to the mathematical concept of linking
number~\cite{ricca2011}.  It is possible to quantify such linking by
means of Gauss
integrals~\cite{Panagiotou_JPA_2010,Panagiotou_PRE_2013,Baiesi_et_al_SciRep_2016},
from which the proposed name of Gaussian
entanglement~\cite{Baiesi_et_al_JPA_2017,Baiesi_SR_2019}.  Also this
kind of intricacy may lead to a slowing down of the folding, as
suggested by the significant correlation between Gaussian entanglement
and folding rates~\cite{Baiesi_et_al_JPA_2017}. Interestingly, the
Gaussian entanglement and the contact order can be combined to improve
the predictions of folding rates~\cite{Baiesi_et_al_JPA_2017}.

Recently, it was discovered that entangled loops (i.e.~looped portions
of a protein with large Gaussian entanglement with another portion of
the same protein) appear in roughly one third of known single domain
proteins~\cite{Baiesi_SR_2019}, a much larger fraction than that of
knotted proteins~\cite{jackson2017}. Moreover, the amino acids at the
closures of entangled loops have a mutual attraction which is on
average, weaker than in the set of all
closures~\cite{Baiesi_SR_2019}. A plausible explanation of the
statistical lack of stable closures for entangled loops is that they
would require a complicate threading by another part of the protein
after their early formation. It is likely better for the folding
dynamics to perform the closure of entangled loops as late as
possible. This hypothesis is corroborated by an asymmetry in the
position of entangled loops with respect to the chain portion they are
entangled with, the thread, such that the latter is found more
frequently on the loop N-terminal side~\cite{Baiesi_SR_2019}. In the
context of cotranslational folding~\cite{waudby2019}, this would imply
that entangled loops are synthesized at the ribosome, and hence
folded, on average later than the thread.

In this work we aim at understanding if the weak closure of entangled
loops can be interpreted, at least in principle, as the result of a
selective pressure that optimizes the folding rate. We do this within
a simple toy model where short protein chains are defined on a
face-centered cubic (fcc) lattice, forming a population with random
initial amino acid sequences that are then subject to an evolutionary
process. We consider two different putative native states sharing
exactly the same ground state energy (for similar sequences) and similar
network topologies. However, one is characterized by a large Gaussian
entanglement (in the form of two concatenated loops), whereas the
second one presents no significant entanglement and is used as a
negative control. This simple model allows a sufficiently quick
repetition of the folding dynamics for many protein copies within a
structure-based approach~\cite{go1983}. At every step of the
evolutionary process, the sequence with the longest folding time is
replaced by another sequence, so that the population evolves toward a
state where the entangled loops are indeed on average more weakly
bound at their closures than other non-entangled loops in the same
structure. This effect is much less enhanced in the negative control
case, when all loops in the native structure are not entangled.
 
\section{Results}

\begin{figure}[t]
  \centering
  \includegraphics[width=0.7\columnwidth]{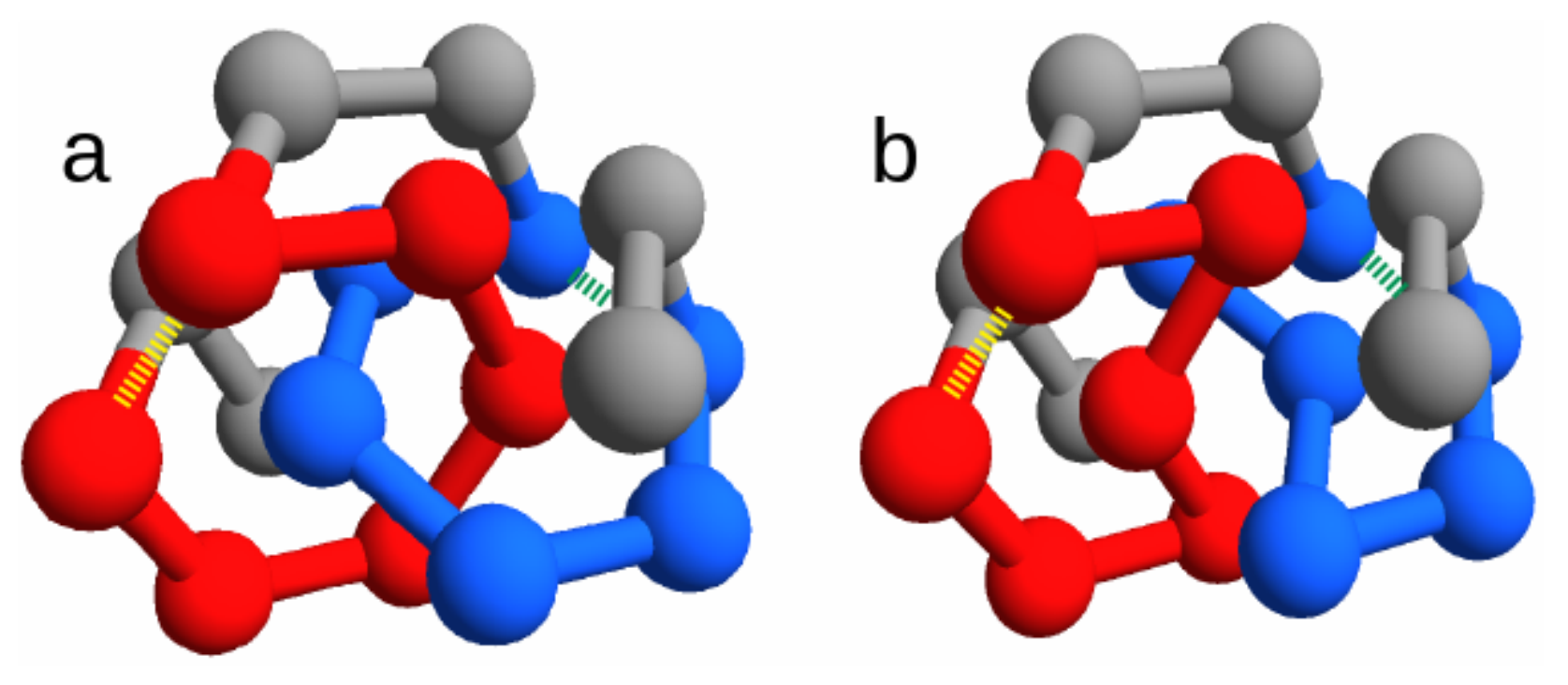}
  \caption{(\textbf{a}) Entangled native state. (\textbf{b})
    Non-entangled native state. The putative ``twin'' native states
    are two self-avoiding walks on the fcc lattice with $N=18$
    sites. The two loops shown in red and blue are concatenated in the
    entangled structure, and not concatenated in the non-entangled
    structure. The amino acids at the ends of each loop form the
    contacts (dashed yellow and green lines) whose energy is studied
    in this work.}
  \label{fig:conf}
\end{figure}

In our toy model for protein chains, we consider a structure-based
energy function with a sequence-dependent energy
(Section~\ref{ssec:protchain} for details). Once a native structure is
chosen to define the energy function, all sequences in the model will
have that structure as a ground state, with a sequence-dependent
ground state energy. In this study, we consider two putative native
states. One state is entangled, with two concatenated loops, whereas
the second ``twin'' state is non-entangled (Figure~\ref{fig:conf}).

Despite the different overall topology, the two twin states display a
very similar contact network topology (compare
Table~\ref{tab:protlink} and Table~\ref{tab:protnolink}). In fact, one
of the two states can be converted into the other by just switching
the spatial positions of two particular amino acids,
so that only a few contacts are rewired. The overall energy, however,
can be kept exactly the same, upon also switching the corresponding amino
acid types
(Section~\ref{ssec:natconf} for details).

We focus on the two energies of the contacts involved in the
closures of the loops, which can be either the concatenated loops
in the entangled native state or 
the corresponding loops in the ``twin'' non-entangled state.
These are represented as dashed lines in Figure~\ref{fig:conf}, joining
amino acid $3$ with $8$ and amino acid $11$ with $16$.
Due to the symmetry of the conformations, the two contacts are in equivalent
positions. Therefore, for any sequence $s$,
we can distinguish the energies $V_1(s) >V_2(s)$ of the contacts with,
respectively, the weaker (higher energy) and the stronger interaction
(more stable due to lower energy).

\begin{table}[tb!]
  \caption{For the entangled protein, pairs $i\div j$ for which $\Delta_{i,j}=1$. Underlined pairs refer to the contacts at the ends of loops discussed in this work. Pairs in bold-face (10 over 35) refer to the contacts that are not present in the non-entangled ``twin'' (Table~\ref{tab:protnolink}).} 
   \label{tab:protlink}
   \small
   \centering 
   \begin{tabular}{lcccccr} 
   $1\div 4$  & $2\div 9$  & ${\bf 4\div 13}$ & $6\div 13$ & ${\bf 6\div 18}$ & $7\div 17$  & $10\div 17$\\
   $1\div 5$  & $2\div 12$ & $5\div 13$ & $6\div 14$ & $7\div 9$  & $7\div 18$  & \underline{$11\div 16$}\\
   $1\div 12$ & ${\bf 2\div 13}$ & $5\div 14$ & ${\bf 6\div 15}$ & $7\div 10$ & ${\bf 8\div 13}$  & $11\div 17$\\
   ${\bf 1\div 13}$ & \underline{$3\div 8$}  & ${\bf 6\div 11}$ & ${\bf 6\div 16}$ & $7\div 12$ & $9\div 12$  & $14\div 18$\\
   $2\div 8$  & ${\bf 3\div 13}$ & $6\div 12$ & ${\bf 6\div 17}$ & $7\div 13$ & $10\div 12$ & $15\div 18$\\
   \end{tabular}

   \caption{For the twin protein without link, pairs $i\div j$ for which $\Delta_{i,j}=1$. Underlined pairs refer to the contacts at the ends of loops discussed in this work. Pairs in bold-face (10 over 35) refer to the contacts that are not present in the entangled structure (Table~\ref{tab:protlink}).} 
   \label{tab:protnolink}
   \small
   \centering 
   \begin{tabular}{lcccccr} 
   $1\div 4$  & $2\div 8$  & ${\bf 4\div 6}$  & $6\div 13$ & $7\div 13$  & $10\div 17$ & ${\bf 13\div 16}$\\
   $1\div 5$  & $2\div 9$  & $5\div 13$ & $6\div 14$ & $7\div 17$  & ${\bf 11\div 13}$ & ${\bf 13\div 17}$\\
   ${\bf 1\div 6}$  & $2\div 12$ & $5\div 14$ & $7\div 9$  & $7\div 18$  & \underline{$11\div 16$} & ${\bf 13\div 18}$\\
   $1\div 12$ & ${\bf 3\div 6}$  & ${\bf 6\div 8}$  & $7\div 10$ & $9\div 12$  & $11\div 17$ & $14\div 18$\\
   ${\bf 2\div 6}$  & \underline{$3\div 8$}  & $6\div 12$ & $7\div 12$ & $10\div 12$ & ${\bf 13\div 15}$ & $15\div 18$\\
   \end{tabular}
\end{table}

\subsection{Concatenated loops slow down the folding of random
  sequences}
\label{ssec:random}
\begin{figure}[tb]
  \centering
    \includegraphics[width=0.8\columnwidth]{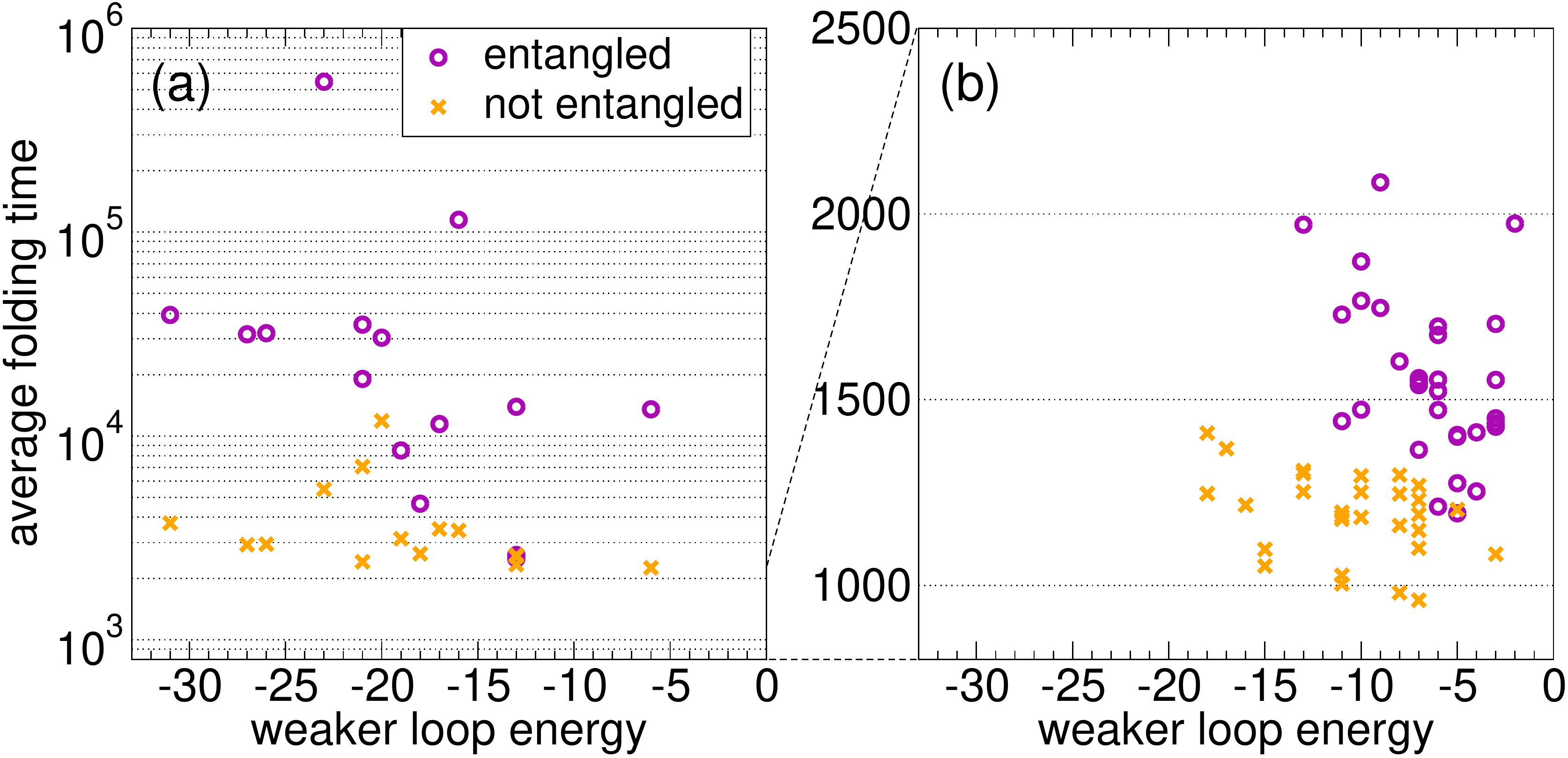}
    \caption{Average folding times into the entangled native state
      (crosses) and into the twin native state without entanglement
      (circles), as a function of the weaker of the contact energies
      involved at the ends of the two loops ($V_1$). (a): $15$
      independent random sequences; (b): fastest proteins after
      $G=1000$ generations of the evolutionary process for $31$
      independent replicas. Evolution leads to a
      dramatic drop in time scales (note the log scale for the random sequences),
      yet the entangled proteins, with respect to their non-entangled twins, 
      still fold more slowly.
      Note also that the evolved entangled proteins have on average more unstable
      energetic closures of the loops.
    }
  \label{fig:tw}
\end{figure}

We begin by comparing the average folding times of random sequences
that fold onto the entangled native state, shown in
Figure~\ref{fig:conf}(a), with those of their twin sequences that fold
onto the twin non-entangled native state, shown in
Figure~\ref{fig:conf}(b), with the same ground state energy. The random
sequences $\{a_1,\ldots,a_p,\ldots,a_q,\ldots,a_N\}$, assigned to the
entangled native state, are sampled with a uniform distribution over
all 20 possible amino acid types. The related twin sequences can then
be defined as $\{a_1,\ldots,a_q,\ldots,a_p,\ldots,a_N\}$. We perform
$n=100$ runs to estimate the average folding times for each
considered sequence (Section~\ref{ssec:fold} for simulation
details). This procedure is repeated for $15$ random sequences and
their twins. Figure~\ref{fig:tw}(a) shows the average folding time as a
function of the weaker energy $V_1(s)$. As expected, the proteins
that fold onto the entangled state need on average much more time than
their twins. For random sequences, in the absence of an evolutionary
process, the folding time is not correlated with the loop closure
energy in both the link and the no-link case. A similar conclusion
can be drawn for the stronger energy $V_2(s)$.

\subsection{Folding time optimization results in slightly lower average
  energies for the concatenated protein structure}
\label{ssec:evres1}
After having established the folding kinetic properties of random
sequences, we now study the outcome of an evolutionary process that
optimizes the average folding time for the resulting sequences
(Section~\ref{ssec:evol} for more details on the simulated
evolutionary process).

For a given choice of the putative native state, the evolutionary
process is simulated for $S=20$  independent replicas. In each replica,
$Z=100$ proteins with random initial sequences are evolved for a total
of $G=1000$ evolutionary steps, or generations. At the end of the
process, the final ensemble of each replica consists of $Z$ optimized
sequences with statistical properties distinguished from the initial
random ones.

We simulate the evolutionary process with either the entangled
conformation (Figure~\ref{fig:conf}(a)), or its non-entangled ``twin''
(Figure~\ref{fig:conf}(b)), chosen as the respective native state. We
focus on the properties of the ``best'' protein $\omega^*_s$, i.e.~the
protein with the lowest average folding time $\tau_s =
\mean{\tau(\omega^*_s)}$, within each system $1\le s\le S$.

The resulting native energy per residue $E$ of such proteins, averaged
on the ensemble of $S$ independent replicas, is slightly lower in the
link case ($\left<E\right>= -22.13 \pm 0.16$) with respect to the
no-link case ($\left<E\right>= -21.76 \pm 0.15$). This energy
difference is significant at the level of $1.7$ standard deviations
($p$-value 0.045 with a one-tailed test); it may be needed to
compensate for the entropy loss caused by the rigidity due to loop
concatenation~\cite{Baiesi_SR_2019}.

\subsection{Folding time optimization promotes weak interactions at
  the end of concatenated loops}
\label{ssec:evres2}

For all proteins with the lowest average folding time, the latter is
shown in Figure~\ref{fig:tw}(b) as a function of the energy $V_1(s)$
of the weakest closure
(a similar pattern is found for $V_2(s)$), for the $S$ fastest proteins
evolved on the entangled native state and for the $S$ fastest proteins
in the no-link case.  Clearly, the former ones on average fold more
slowly than the latter ones. Most importantly, proteins evolved on the
entangled native state are characterized by higher $V_1$ values,
i.e.~the closures of concatenated loops are less stable as a result of
the evolutionary process, even if their overall native energy is lower
(Section~\ref{ssec:evres1}).

On average, over the $S$ independently replicated evolutionary
processes, we find $\overline \tau =\frac 1 S \sum_s \tau(s) = 1560
\pm 40$, $\overline V_1=\frac 1 S \sum_s V_1(s) = -6.5 \pm 0.5$,
$\overline V_2=\frac 1 S \sum_s V_2(s) = -11.4 \pm 0.7$, in the
presence of an entangled native state. In the no-link case, we get
instead significantly lower values for all quantities (at a level of,
respectively, $8.3$, $5.1$, $5.4$ standard deviations in the different
cases): $\overline \tau = 1190 \pm 20$, $\overline V_1 = -10.9 \pm
0.7$, $\overline V_2 = -18.4 \pm 1.1$.

\begin{figure}[tb]
  \centering
  \includegraphics[width=0.8\columnwidth]{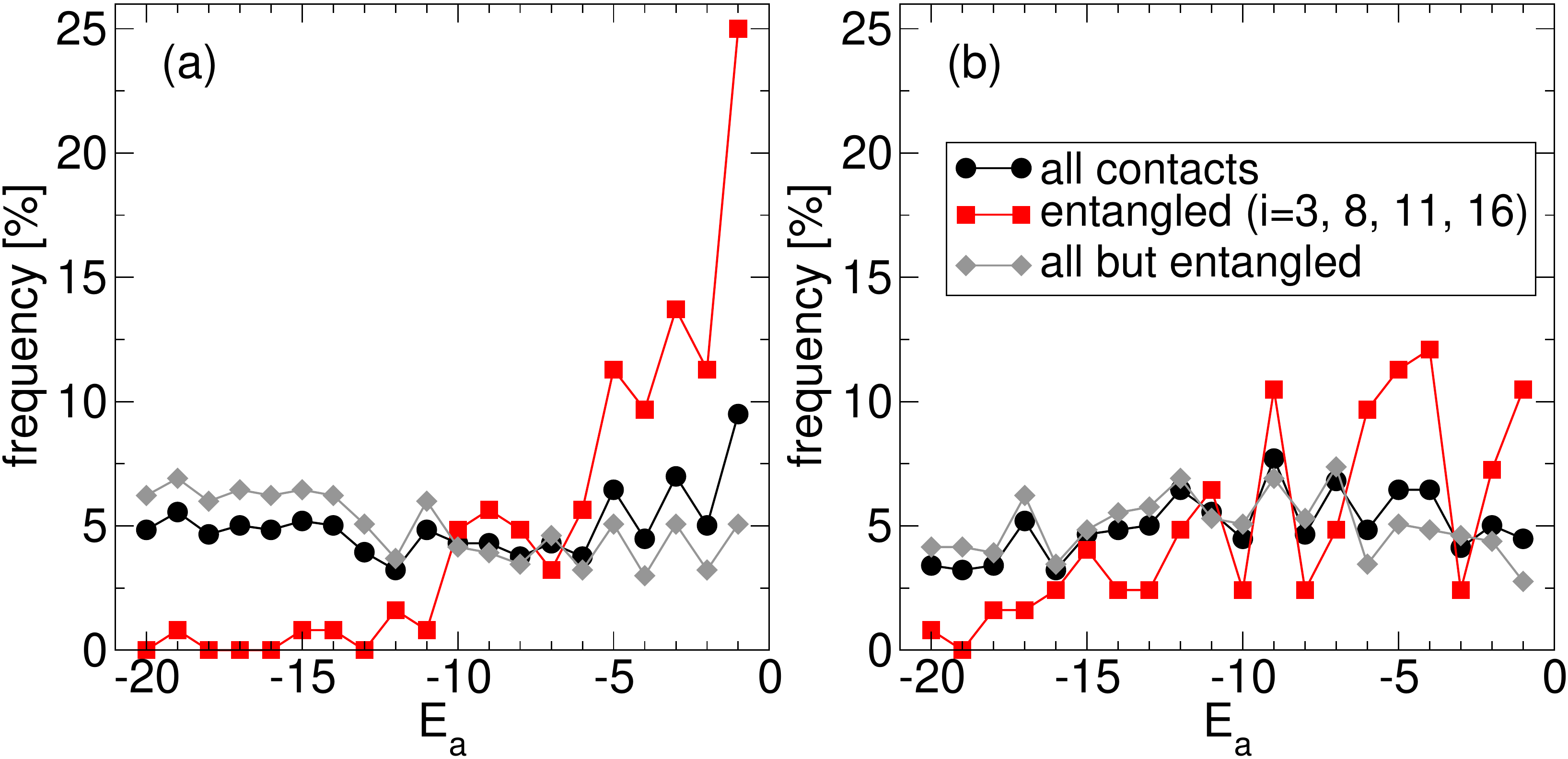}
  \caption{Frequency of amino acid single potentials for (a) protein with entangled loops and (b) its twin without entanglement, collecting the statistics of proteins at the end of the evolutionary process. The three curves show the frequency regardless of the position along the chain, the frequency at the $4$ sites closing the loops, and the frequency in the complementary set of sites.}
  \label{fig:ev}
\end{figure}

\subsection{Folding time optimization promotes hydrophilic residues at
  the end of concatenated loops}
\label{ssec:evres3}

Consistently with the results shown in the previous Section, our
findings reveal also that folding time optimization on the entangled
native structure leads to sequences where concatenated loops are
closed by amino acids that are less hydrophobic than the average
one. This is made apparent in Figure~\ref{fig:ev}(a), where we focus
again on the ensemble of the $S$ fastest proteins found at the end of
the corresponding independent replicas of the evolutionary process. We
plot the frequency observed for each amino acid type $a$ (the
corresponding hydrophobicities are negative integers, $-20\le
E_a\le-1$ in our toy model, Section~\ref{ssec:protchain}): (i)
regardless of its position along the chain, (ii) at one of the $4$
sites at the ends of the two concatenated loops (dashed lines in
Figure~\ref{fig:conf}(a)), and (iii) at the complementary $N-4$ sites.
Case (i) and (iii) show that, on average, all amino acids are equally
frequent, consistently with the sampling of amino acid types used in
the evolutionary process. Note that, in principle, one might have
expected an overall bias towards more hydrophobic residues, based on
the naive expectation that the stronger the interactions the faster
they form, but this is not the case. The slight difference between (i)
and (iii) is due to the inclusion in (i) of the statistics (ii) of the
$4$ special sites, which shows a significant departure from the flat
profile. Indeed, due to the evolutionary pressure promoting fast
folding, the most hydrophobic residues are selected against at the end
of concatenated loops (amino acids with $E_a<-10$ are nearly absent),
whereas hydrophilic ones are instead found much more frequently (the
distribution has a large peak in correspondence of the more
hydrophilic amino acid with $E_a=-1$).

It is important to check that this trend is actually due to the
presence of two concatenated loops and not simply to the overall
arrangement of the native structure. The corresponding frequencies
observed for the amino acid types in the no-link case, where the $4$
``twin'' sites in the non-entangled state (connected by dashed lines
in Figure~\ref{fig:conf}(b)) are either singled out or excluded, are
shown in Figure~\ref{fig:ev}(b). The residue type selection observed in
the link case is much stronger than in the no-link case. Nevertheless,
a similar, albeit much slighter, trend is present also in the latter
case, with hydrophilic residues found more frequently than hydrophobic
ones at the $4$ special sites. The most hydrophobic residues can
anyhow still be found in a significant amount.

\section{Discussion}

Within a toy model for short protein chains, we simulated an
evolutionary process where folding time is optimized for a given
native structure.  Coarse-grained structure-based models are commonly used
to study folding kinetics, in particular in the context of knotted and
entangled proteins~\cite{Cieplak2015,Zhao2018,perego2019a,perego2019r}.

In order to better understand the role of entanglement within the
protein, we considered two similar native structures that are related
by a subtle rewiring of few chain bonds and thereby differ only for
the presence of a pair of concatenated loops in just one of them
(Figure~\ref{fig:conf}). For any given sequence that folds onto the
entangled native structure, a ``twin'' sequence can be obtained by
switching two amino acid types, having the same ground state energy
onto the non-entangled native structure.

Despite the simplicity of our approach, the results reproduce sequence
patterns related to the presence of entangled motifs that were
detected by analyzing single-domain protein
structures~\cite{Baiesi_SR_2019}. Namely, the evolutionary process
leads to optimized sequences whose amino acids are enriched in
hydrophilic residue types at the end of concatenated loops, when the
latter are present in the native structure (Figure~\ref{fig:ev}).  The
results obtained within the toy model thus corroborate the hypothesis
that the need to perform a fast and smooth folding process has
selected amino acid sequences where some degree of frustration, in the
form of unfavorable amino acid pair stability, is allowed. This
energetic frustration is localized at the ends of
concatenated/entangled loops, allowing to overcome the topological
frustration implied by the presence of entangled motifs, in keeping
with the principle of minimal
frustration~\cite{bryngelson1987,frauenfelder1991}.

The crucial role of loop concatenation is benchmarked against the
results obtained for the non-entangled native structure. A residual
selection of hydrophilic residues is observed also in this case,
hinting that the evolution of weak interactions to allow fast folding
may be a feature non restricted to the closures of entangled loops. At
any rate, the observed enrichment in hydrophilic residues is markedly
weaker than for the entangled structure (Figure~\ref{fig:ev}).  At the
same time, a bias, if any, is observed instead towards lower overall
native energies of the evolved sequences for the entangled native
structure (Section~\ref{ssec:evres1}).

In general, folding is on average much slower in the presence of
concatenated loops, as expected. This holds true when comparing the
ensemble of sequences evolved independently on the two native
structures (Figure~\ref{fig:tw}(b)), and also when comparing the
folding of random sequences onto the entangled native structure with
their ``twin'' sequences on the non-entangled structure
(Figure~\ref{fig:tw}(a)).

To sum up, our results support the following picture: given a specific
three dimensional arrangement of residues in the native structure, if
evolution selects sequences enhancing the folding rate, a crucial
byproduct is the removal of strong stabilizing interactions at the
ends of loops, in particular at the end of concatenated loops that
presumably need to be formed in the latter stages of the folding
process.

\section{Materials and Methods}

\subsection{Protein chain model}
\label{ssec:protchain}

We model a protein as a $N$-site self-avoiding walk on the fcc
lattice. Each residue $i\in\{1,2,\ldots,N\}$ carries an amino acid
type $a_i$ with hydrophobicity quantified from $E_a=-a$ for
$a\in\{1,2,\ldots,19,20\}$. Hydrophilic amino acids have $E_a$ closer
to zero and thus form weaker binding with other residues, as described
next.

Non consecutive residues $i,j$ ($|i-j|>1$) form a contact if they are
nearest neighbors on the fcc lattice. We follow a structure-based
Go-like approach \cite{go1983} and assign an energy to any such
contact as

\begin{equation}
V_{i,j} = E_{a_i,a_j} \cdot \Delta_{i,j} =  \left(E_{a_i} + E_{a_j}\right) \Delta_{i,j}
\label{eq:add}
\end{equation}

Here $\Delta_{i,j}=0,1$ is the native connectivity matrix (the contact
map) in the Go-like model, namely $\Delta_{i,j}=1$ only if the contact
$i\div j$ is present in the native state. The energetic contribution
$E_{a_i,a_j}\equiv E_{a_i} + E_{a_j}$ is the simplest linear
combination of the amino acid hydrophobicities. Note that all contact
energies are attractive in our model, so that any given structure with
a non zero connectivity matrix is the ground state for all sequence
choices. The overall energy for a given chain configuration $\Gamma$
with amino acid sequence $\left\{a_i\right\}_{i=1}^N$ is obtained by
summing over all nearest-neighbor interactions on the lattice among
non consecutive residues along the chain:
\begin{equation}
E\left(\Gamma,\left\{a_i\right\}\right) = \sum_{j>i+1}V_{i,j} = \sum_{j>i+1}\left(E_{a_i} + E_{a_j}\right) \Delta_{i,j}
\label{eq:ene}
\end{equation}

Note that the simple additive form Eq.~(\ref{eq:add}) of the pairwise
interaction potential was shown to capture much of the statistical
variability of the interaction parameters derived by Miyazawa and
Jernigan in a knowledge-based approach~\cite{li1997}.

\subsection{Native conformations}
\label{ssec:natconf}

We consider two alternative native state configurations for short
self-avoiding chains ($N=18$), shown in Figure~\ref{fig:conf}. Both
structures are the ground states for any sequence choice, in the
corresponding structure-based models defined by Eq.~(\ref{eq:ene}). They
are non-degenerate only up to mirror images, since we do not consider
terms that break the chiral symmetry. The list of their contacts in the
native conformation is presented in Table~\ref{tab:protlink} for the
entangled protein and in Table~\ref{tab:protnolink} for the twin
without entanglement.

The choice of short chains allows fast simulations, at the same time
leaving the possibility for a non trivial topological structure that
exhibit a pair of concatenated loops. We choose this entangled
structure, $\Gamma_l$, as the main object of our study (see
Figure~\ref{fig:conf}(a)).

The fcc lattice allows to build an almost identical non-entangled
structure, $\Gamma_{nl}$ (through the text we called it the ``twin'' structure), where no
pair of loops is concatenated (Figure~\ref{fig:conf}(b)). The twin
non-entangled structure $\Gamma_{nl}$ can be obtained by just
switching the spatial positions ($\vec{r}_p \leftrightarrow \vec{r}_q$)
of the two residues $p=6$ and $q=13$ in the entangled structure
$\Gamma_l$. Note that this entails a rewiring of chain connectivity,
consistently with the change in the overall topology. With a simple
notation, the two twin structures will be hereafter labeled as ``link''
and ``no link'' in the figures.

If, in addition, one performs a similar switch ($a_p \leftrightarrow
a_q$) between the corresponding amino acid types in the sequence, all
amino acid types turn out to be kept in the same three-dimensional
positions. However, the values of 4 contact energies are modified
under the combined structural and sequence switches, in correspondence
of the interacting pairs affected by chain rewiring.

Nevertheless, the simple form Eq.~(\ref{eq:add}) for the amino acid
pairwise interaction potential ensures that, the sequence
$\left\{a_i\right\}$, with ground state energy
$E\left(\Gamma_l,\left\{a_i\right\}\right)$ on the entangled
structure, will have {\it exactly} the same ground state energy as the
switched ``twin'' sequence $\left\{a'_i\right\}$ on the non-entangled twin
structure:
$E\left(\Gamma_l,\left\{a_i\right\}\right)=E\left(\Gamma_{nl},\left\{a'_i\right\}\right)$.
Thanks to the above property, the non-entangled structure does
not have any energetic advantage with respect to the
entangled structure $\Gamma_l$ during the folding of 
selected sequences.

Finally, we note that there is an additional symmetry for both
structures, because the inversion of the chain direction
($i\leftrightarrow N-i+1$) produces its mirror image. Accordingly,
the detected evolutionary signals will not depend on the asymmetry in
the location of the concatenated loops along the chain.

\subsection{Folding simulations}
\label{ssec:fold}

\begin{figure}[tb]
  \centering
    \includegraphics[width=0.999\columnwidth]{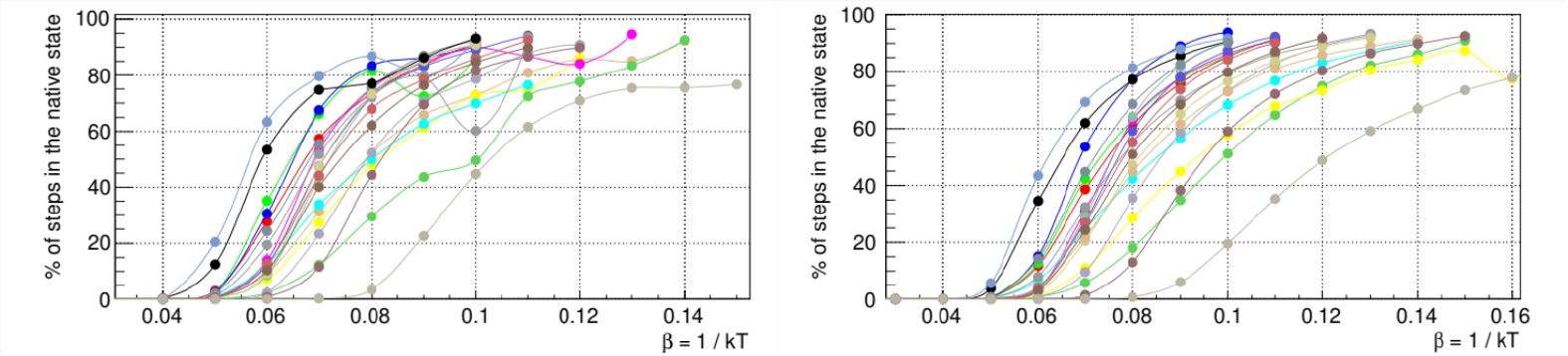}
    \caption{Fraction of configurations in the native states as a
      function of the inverse temperature $\beta=1/k_B T$, for
      proteins with entanglement (left) and without entanglement
      (right). Each curve is for a given random sequence. The mean
      folding inverse temperature is found by averaging the points
      where the curves cross the $50\%$. The inverse temperature used
      in the folding simulations within the evolutionary process is
      fixed to this value.}
  \label{fig:temp}
\end{figure}

The time trajectory of each protein in the conformation space is
initialized from a random high-temperature configuration. The
temperature is switched at time $t=0$ to a value
$T=1/\beta=1/0.071\simeq 14.1$ (with units in $k_B=1$) that was
determined by averaging the folding thermodynamics properties of
random sequences (Figure~\ref{fig:temp}). This temperature leads to
the eventual folding of the protein, a stochastic process that we
simulate $n=100$ times. Each realization $1\le \alpha\le n$ for a
given protein $\omega$ takes place in a time $\tau_\alpha(\omega)$ and
an average folding time is then evaluated as $\mean{\tau(\omega)} =
\frac 1 n \sum_{\alpha=1}^n \tau_\alpha(\omega)$.

\begin{figure}[tb]
  \centering
    \includegraphics[width=0.8\columnwidth]{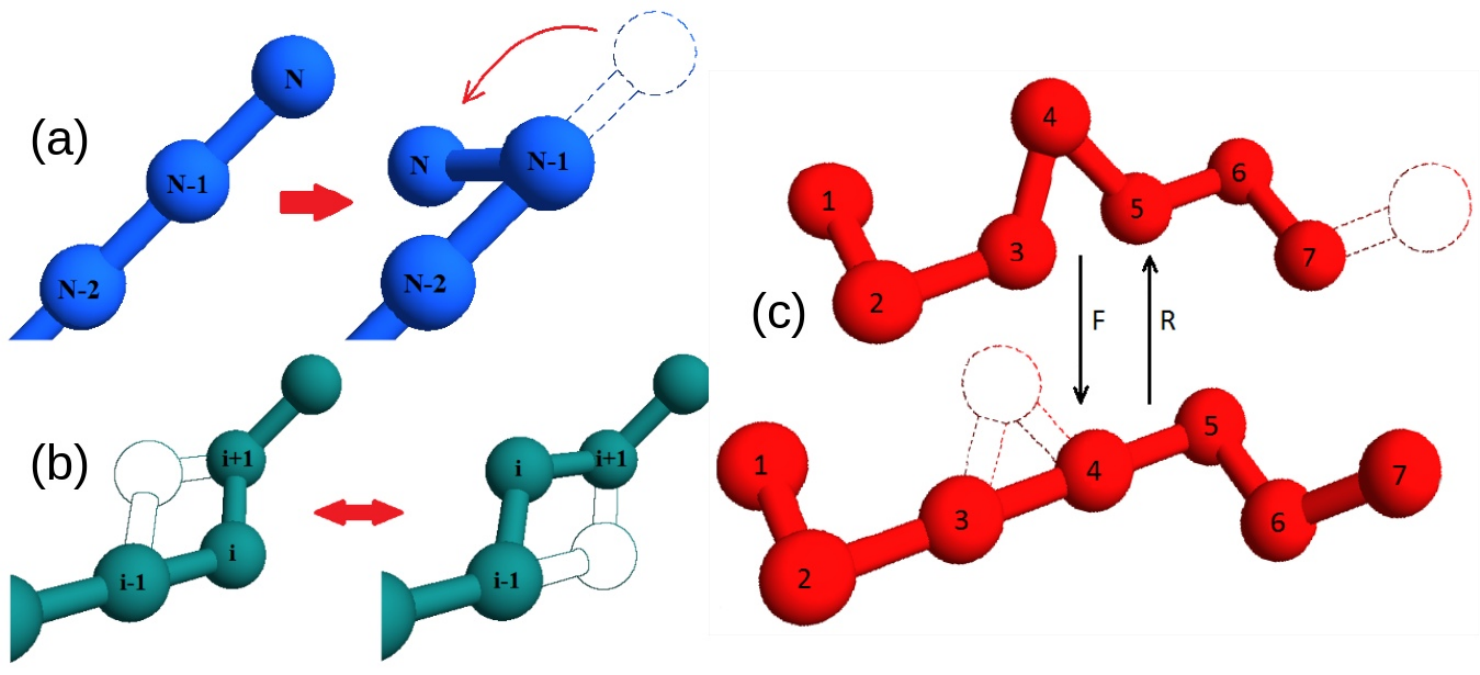}
    \caption{Illustration of the types of moves used in the
      simulations: (a) end-flip (frequency $P=2/19$, the figure
      illustrates the random choice at the end with site $N$), (b)
      crankshaft or internal flip ($P=16/19$), (c) reptation
      ($P=1/19$). In reptation, to satisfy detailed balance, the
      frequency of attempted moves in the two directions satisfy
      $W^R=3 W^F$ (``F'' and ``R'' following the notation in the
      figure) because there are $12$ possible points for the added end
      vs only $4$ internal sites for the added corner. The internal
      flip is in one out of the $4$ possible sites if the corner is of
      $60^o$ or $90^o$, while only two sites are allowed when the
      corner is $120^o$. All attempted moves are then validated with
      self-avoidance constraints and eventually accepted with a
      Metropolis algorithm.}
  \label{fig:moves}
\end{figure}

Protein time dynamics is simulated thanks to a set of Monte Carlo
moves, including both local (crankshaft and end-flip) moves and a
global sliding/reptation move, which have been carefully implemented
to satisfy detailed balance (Figure~\ref{fig:moves}).

Time is measured in units of Monte Carlo sweeps, one sweep containing
a fixed amount $M=N+1$ of Monte Carlo moves. Local crankshaft and end-flip moves
are selected at random with probability $N/M$ and the global reptation
move is chosen with probability $1/M$ to meet the intuition that a
global rearrangement of the backbone is less likely to occur than a
random local displacement. We include the global sliding move to endow
the dynamics with the chance of threading a portion of the backbone
through an already formed loop.

\subsection{Evolutionary process}
\label{ssec:evol}

$Z=100$ protein sequences, each representing an organism, are involved
in the evolutionary process. All sequences are assumed to fold to a
fixed native structure according to the Go-like model defined in
Eq.~(\ref{eq:ene}). The native structure can be chosen as either the
entangled structure in Figure~\ref{fig:conf}(a), or the twin
non-entangled structure in Figure~\ref{fig:conf}(b). At beginning of the
process, protein sequences are initialized by choosing randomly each
amino acid with a uniform probability across all 20 possible
types. During the process, we evaluate the average folding times for
all proteins as described in Section~\ref{ssec:fold}.  The longest
folding time is associated with the lowest fitness of the organism
hosting that protein. For simplicity, this just leads to its
extinction, and its position in the niche is occupied by another
element. For its replacement we follow this procedure: with
probability $p_r=1/3$ a completely new random sequence of amino acids,
sampled with uniform probability as above, is assigned to the newborn;
otherwise the sequence of the faster folder is copied with partial
fidelity, i.e.~amino acids are kept the same with probability
$p_c=0.9$ and uniformly sampled at random otherwise. The protocol for
generating the new sequence thus includes the priority gained by the
organism with the best fitness (yet allowing mutations of its genome)
to populate the empty slot, but also the possibility of a random
entrance of brand new organisms in the empty niche.

\end{document}